\documentclass[%
reprint,
superscriptaddress,
amsmath,amssymb,
aps,
pra
]{revtex4-1}
\usepackage{graphicx}% Include figure files
\usepackage{dcolumn}
\usepackage{bm}
\usepackage{physics}
\usepackage{braket}
\begin{document}
\title{The nature of the Lamb shift in weakly-anharmonic atoms:\\ from normal mode splitting to quantum fluctuations}

\author{Mario F. Gely}
\affiliation{%
Kavli Institute of NanoScience, Delft University of Technology,\\
PO Box 5046, 2600 GA, Delft, The Netherlands.
}
\author{Gary A. Steele}

\affiliation{%
Kavli Institute of NanoScience, Delft University of Technology,\\
PO Box 5046, 2600 GA, Delft, The Netherlands.
}%

\author{Daniel Bothner}%
 \affiliation{%
Kavli Institute of NanoScience, Delft University of Technology,\\
PO Box 5046, 2600 GA, Delft, The Netherlands.
}%

\date{\today}

\begin{abstract}
When a two level system (TLS) is coupled to an electromagnetic resonator, its transition frequency changes in response to the quantum vacuum fluctuations of the electromagnetic field, a phenomenon known as the Lamb shift. 
Remarkably, by replacing the TLS by a harmonic oscillator, normal mode splitting leads to a quantitatively similar shift, without taking quantum fluctuations into account. 
In a weakly-anharmonic system, lying in between the harmonic oscillator and a TLS, the origins of such shifts can be unclear. 
An example of this is the dispersive shift of a transmon qubit in circuit quantum electrodynamics (QED).
Although often referred to as a Lamb shift, the dispersive shift observed in spectroscopy in circuit QED could contain a significant contribution from normal-mode splitting that is not driven by quantum fluctuations, raising the question: how much of this shift is quantum in origin?
Here, we treat normal-mode splitting separately from shifts induced by quantum vacuum fluctuations in the Hamiltonian of a weakly-anharmonic system, providing a framework for understanding the extent to which observed frequency shifts can be attributed to quantum fluctuations. 
\end{abstract}
\maketitle
% \twocolumngrid

\section{Introduction}
% \subsection{There's a problem with the origin of the Lamb shift... }

Quantum theory predicts that vacuum is never at rest. 
On average, the electromagnetic field of vacuum has no amplitude, but quantum vacuum fluctuations impose a fundamental uncertainty in its value. 
This is notably captured in the ground-state energy of a harmonic oscillator (HO) $\hbar\omega_r/2$. 
When an atom couples off-resonantly to an electromagnetic mode, equivalent to a HO, the quantum vacuum fluctuations of the mode shift the transition frequencies between states of the atom~\cite{lamb_fine_1947}. 
This effect is called the Lamb shift. 
If the atom can be modeled as a two level system (TLS), this interaction is described in the rotating wave approximation (RWA) by the Jaynes-Cummings Hamiltonian~\cite{jaynes_comparison_1963}. 
The so-called Lamb shift is then given by $-g^2/\Delta$ in the dispersive regime $g\ll |\Delta|$ where $g$ is the coupling strength and $\Delta = \omega_r-\omega_a$ is the frequency detuning between the mode ($\omega_r$) and atom ($\omega_a$). 

If one replaces the TLS with a HO, a similar effect occurs from normal-mode splitting, where in the dispersive regime, each oscillator acquires a frequency shift due to the presence of the other oscillator. 
This similarity is not only qualitative: in the RWA parameter regime, a classical calculation of the normal mode splitting of two HOs also predicts this shift to be $-g^2/\Delta$.
A quantum calculation for two HOs will also give the same result: this shift for HOs is not influenced by the presence of quantum fluctuations.
Extending this further, one can replace the TLS atom with a weakly-anharmonic oscillator, such as a transmon qubit in circuit QED.
In experiments in circuit QED, a shift of $-g^2/\Delta$ was also observed, has been attributed to being induced by vacuum fluctuations, and is commonly referred to as the Lamb shift~\cite{fragner_resolving_2008}. 
However, normal mode splitting of two HOs, which includes no effect of quantum fluctuations, also leads to a shift of the same size.
This then raises the following question: how much of the dispersive shift in weakly-anharmonic atoms arises from quantum fluctuations? 
Or equivalently, how much of this shift persists if quantum fluctuations are neglected?

Here, we derive analytical expressions for the quantum fluctuation contribution to the dispersive shift of weakly-anharmonic atoms. 
We find that for a weakly-anharmonic atom coupled dispersively to a harmonic oscillator, two distinct shifts occur; one is a quantum effect due to vacuum fluctuations, another arises from normal-mode splitting. 
To illustrate the described physics, this work focuses on the transmon qubit~\cite{koch_charge-insensitive_2007} coupled to a $LC$-circuit.
We follow the approach of transforming the Hamiltonian to its normal-mode basis~\cite{nigg_black-box_2012} and treating anharmonicity as a perturbation.
By performing calculations analytically, we gain insight into the origin of different frequency shifts, and reach accurate approximations of their magnitude, extending expressions previously derived~\cite{koch_charge-insensitive_2007} to regimes of large detuning. 
Our expression of the AC stark shift decreases with the square of the frequency of a coupled mode, which notably places strong limitations on the coupling of low frequency mechanical elements to these type of qubits~\cite{pirkkalainen_hybrid_2013}.

\begin{figure*}[t!]
\includegraphics[width=0.92\textwidth]{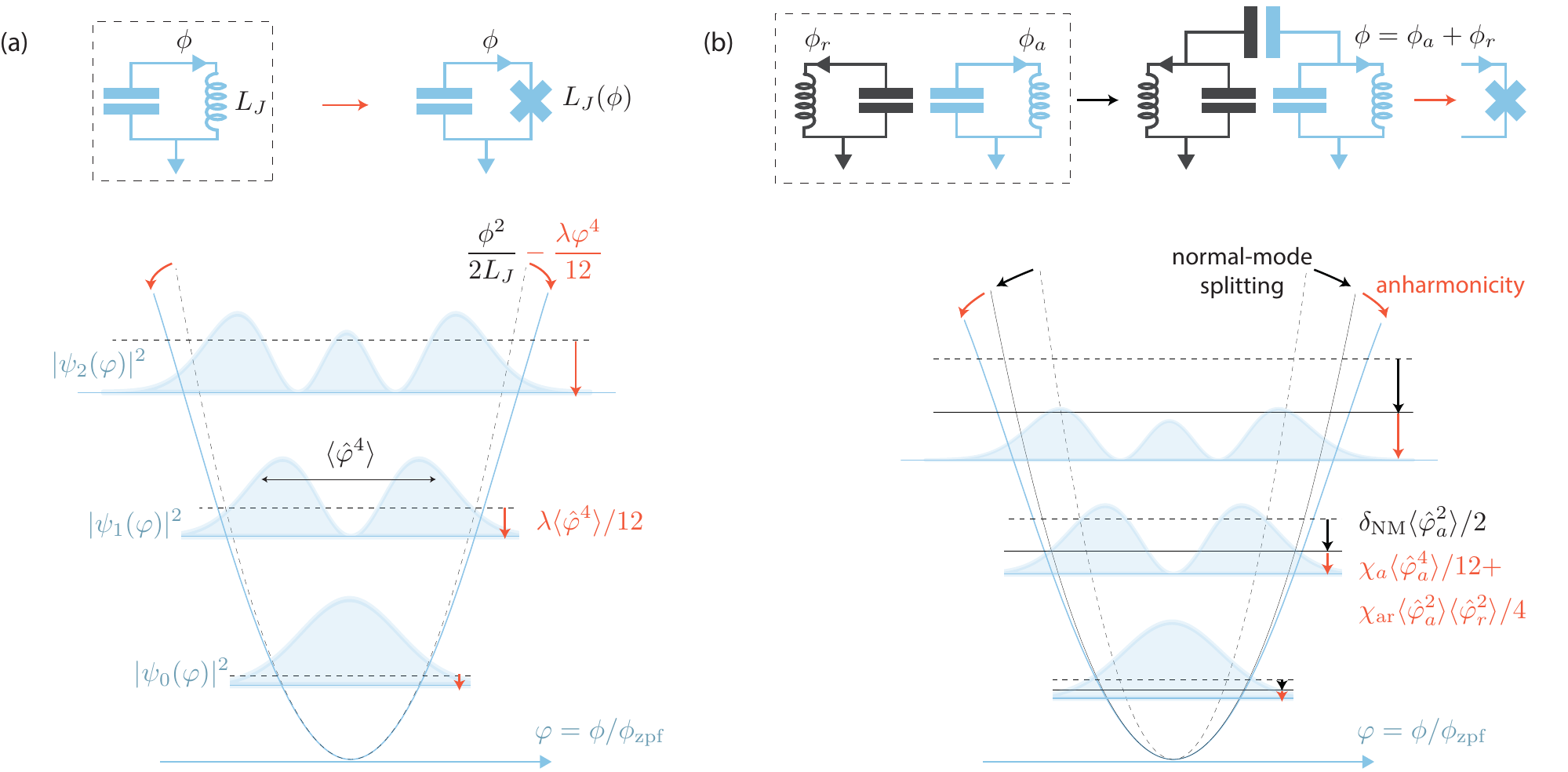}
\caption{The origin of different energy shifts in a weakly-anharmonic atom. 
(a) Replacing the linear inductance of an $LC$ oscillator with a Josephson junction (JJ) results in a weakly-anharmonic artificial atom.
To first order, the energy level $n$ is shifted proportionally to $\bra{n}\hat{\phi}^4\ket{n}$, where $\ket{n}$ is a Fock state of the harmonic system.
(b) Two coupled harmonic oscillators undergo normal-mode splitting, resulting in a frequency shift $\delta_\text{NM}$.
The flux traversing one of the inductances $\phi$, is then composed of the flux from both normal mode oscillations $\phi = \phi_a+\phi_r$.
Replacing an inductor with a JJ leads to the same shift as in the isolated atom $\chi_a\langle\hat{\phi}_a^4\rangle$, but also to a shift due to quantum fluctuations of the coupled oscillator $\chi_{ar}\langle\hat{\phi}_a^2\rangle \langle\hat{\phi}_r^2\rangle$.}
\label{fig:fig1}
\end{figure*}

\section{Weak anharmonicity: case of the transmon qubit}
% \label{sec:transmon}

We define a weakly-anharmonic atom as a harmonic oscillator with a small quartic potential
\begin{equation}
	\hat{H}/\hbar = \underbrace{\omega_a (\hat{a}^\dagger\hat{a}+\frac{1}{2})}_{\hat{H}_\text{HO}}\underbrace{-\frac{\lambda}{12} \left(\hat{a}+\hat{a}^\dagger\right)^4}_{\hat{H}_\text{anh}}\ ,
	\label{eq:hamiltonian_a}
\end{equation}
where $\hat{a}$ is the annihilation operator for excitations in the atom, $\omega_a$ the atomic frequency and $\lambda$ the anharmonicity.
In the limit $\lambda \ll \omega_a$, corrections to the eigen-energies of $\hat{H}_\text{HO}$ due to anharmonicity are to first order equal to $-(\lambda/12) \bra{n}(\hat{a}+\hat{a}^\dagger)^4\ket{n}$, with $\ket{n}$ a number state.
We can expand $(\hat{a}+\hat{a}^\dagger)^4$ and only consider terms that preserve the number of excitations $n$, since only they will give a non-zero contribution to the first-order correction
\begin{equation}
\hat{H}_\text{anh}/\hbar \simeq -\frac{\lambda}{2}\left(\left(\hat{a}^\dagger\hat{a}\right)^2 +\hat{a}^\dagger\hat{a}+ \frac{1}{2}\right)\ ,
\label{eq:levels}
\end{equation}
leading to energy levels
\begin{equation}
E_n/\hbar \simeq (\omega_a-\lambda)\left(n+\frac{1}{2}\right)-\lambda\left(\frac{n^2}{2}-\frac{n}{2}-\frac{1}{4}\right)\ .
\label{eq:levels}
\end{equation}
If we write the transition frequencies of the atom $E_n-E_{n-1}=\hbar\omega_a-n\hbar\lambda$, the weakly-anharmonic level structure shown in Fig.~\ref{fig:fig1}(a) becomes apparent.

One implementation of this Hamiltonian is the transmon qubit~\cite{koch_charge-insensitive_2007}.
In addition to being described by the simple electrical circuit of Fig.~\ref{fig:fig1}(a), this system is highly relevant in many experimental endeavors~\cite{gu2017microwave}, from fundamental experiments in quantum optics~\cite{schuster_resolving_2007,bishop_nonlinear_2009,bosman2017approaching,bosman_multi-mode_2017,kirchmair2013observation}, to quantum simulations~\cite{langford_experimentally_2016} or quantum computing~\cite{takita_demonstration_2016,kelly_state_2015,riste_detecting_2015}.
It is constructed from an $LC$ oscillator where the inductor is replaced by the non-linear inductance $L_J\left(I\right)$ of a Josephson junction (JJ).
The transmon is weakly-anharmonic if its zero-point fluctuations in current are much smaller than the junctions critical current $I_c$.
The current $I$ traversing the JJ when only a few excitations populate the circuit is then much smaller than $I_c$ and $L_J\left(I\right)\simeq L_J\left(1+I^2/2I_c^2\right)$.
Intuitively, the expectation value of the current squared $\langle I^2 \rangle$, on which the inductance depends, will increase with the number of excitations in the circuit.
So with increasing number of excitations $n$ in the circuit, the effective inductance of the circuit increases and the energy of each photon number state $E_n$ will tend to decrease with respect to the harmonic case.
For a rigorous quantum description of the system, the flux $\phi(t) = \int^t_{-\infty}V(t')dt'$, where $V$ is the voltage across the JJ, is a more practical variable to use than current~\cite{vool2017introduction}. 
Note that for a linear inductance $L$, the flux $\phi$ is proportional to the current $I$ traversing the inductor $\phi=L I$. 
Using the conjugate variables of flux and charge the Hamiltonian of Eq.~(\ref{eq:hamiltonian_a}) can be shown to describe the transmon~\cite{koch_charge-insensitive_2007}.
The anharmonicity is given by the charging energy $\hbar\lambda = e^2/2C$, the atomic frequency by $\omega_a = 1/\sqrt{L_JC}$ and the flux relates to the annihilation operator through $\hat{\phi} = \phi_\text{zpf}(\hat{a}+\hat{a}^\dagger)$, where the zero-point fluctuations in flux are given by $\phi_\text{zpf} = \sqrt{\hbar\sqrt{L_J/C}/2}$.
We can recover the intuition gained by describing the system with currents by plotting the eigen-states in the normalized flux basis $\hat{\varphi} = \hat{\phi}/\phi_\text{zpf}$ of the harmonic oscillator in Fig.~\ref{fig:fig1}(a). 
The fluctuations in flux increases with the excitation number, hence the expectation value of the fourth-power of the flux $\langle \hat{\phi}^4\rangle\propto\langle \hat{H}_\text{anh}\rangle$ will increase.
The energy of each eigen-state will then decrease, deviating from a harmonic level structure.

\begin{figure*}[t!]
\includegraphics[width=0.85\textwidth]{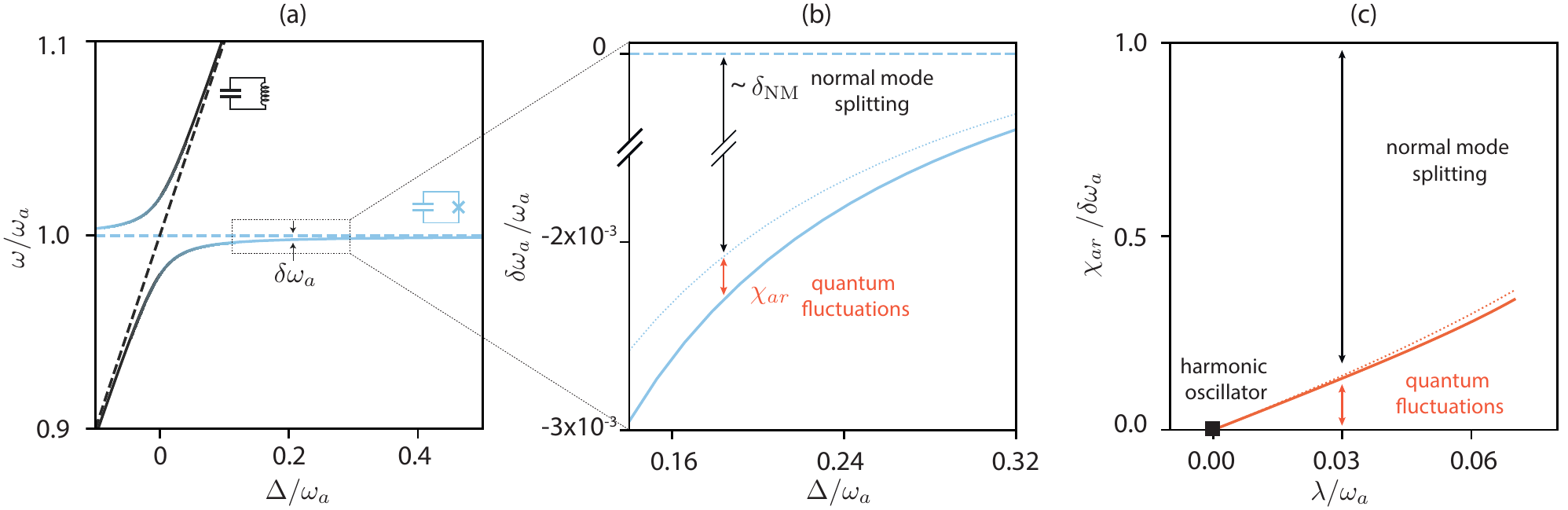}
\caption{Fraction of the atomic energy shift due to quantum vacuum fluctuations. 
(a) Dressed frequency of the ground-to-first-excited state transitions of the harmonic oscillator (black) and atom (blue) as a function of detuning $\Delta = \omega_r-\omega_a$.
Bare frequencies ($g=0$) are shown as dashed lines,
We fixed $\lambda/\omega_a=0.01$ and $g/\omega_a=0.02$.
(b) Total frequency shift $\delta\omega_a$ of the atom, decomposed into its two main components: normal-mode splitting $\delta_\text{NM}$ and a shift resulting from vacuum fluctuations $\chi_{ar}$.
Coupling also changes the anharmonicity $\chi_a$, this results in a small shift absorbed here in $\delta_\text{NM}$.
(c) The vacuum-fluctuations-induced shift $\chi_{ar}$ as a fraction of the total frequency shift of the atom  $\delta \omega_a$ for increasing anharmonicity $\lambda$ and fixed detuning $\Delta = \omega_a/4$.
For a TLS, all of the energy shift arises from quantum fluctuations, $\chi_{ar}/\delta\omega_a = 1$.
In all panels, the dotted lines are computed from Eqs.~(\ref{eq:shifts_beyond}), full lines correspond to a numerical diagonalization of Eq.~(\ref{eq:hamiltonian_a_b_int_anh}). In (c), $\chi_{ar}$ is computed from numerics as half the shift resulting from adding a photon in the oscillator.}
\label{fig:fig2}
\end{figure*}

\section{Coupled harmonic and anharmonic oscillator}
\subsection{Normal-mode splitting and quantum-fluctuation-induced shifts}

We now study the effect of coupling a harmonic oscillator to the atom. 
When an $LC$ oscillator is connected capacitively to a transmon (see Fig.~\ref{fig:fig1}(b)), circuit quantization~\cite{vool2017introduction} leads to the Hamiltonian
\begin{align}
\begin{split}
	\hat{H}/\hbar &= (\omega_a+\lambda) \hat{a}^\dagger\hat{a}-\frac{\lambda}{12} \left(\hat{a}+\hat{a}^\dagger\right)^4\\
	&+\omega_r \hat{b}^\dagger\hat{b}+g \left(\hat{a}-\hat{a}^\dagger\right)\left(\hat{b}-\hat{b}^\dagger\right)\ .
	\label{eq:hamiltonian_a_b_int_anh}
\end{split}
\end{align}
Here $\hat{b}$ is the annihilation operator for photons in the resonator, $\omega_r$ its frequency and $g$ the coupling strength.
Compared to the Hamiltonian of Eq.~(\ref{eq:hamiltonian_a}), we replaced the frequency $\omega_a$ scaling the atomic number operator with $\omega_a+\lambda$.
Doing so will ensure that $\omega_a$ corresponds to the frequency of the first atomic transition, independent of the anharmonicity $\lambda$, as proven by Eq.~(\ref{eq:levels}).
We also omitted the ground-state energies $\hbar\omega_r/2$ and $\hbar(\omega_a+\lambda)/2$ in this Hamiltonian; even though vacuum fluctuations are at the origin of these omitted terms, their presence plays no role in calculating the transition frequencies of the system.
To describe the dispersive regime $g\ll|\Delta|$ of this interaction, we first move to the normal-mode basis, as described in App.~1.
We introduce normal-mode frequencies $\bar{\omega}_{r}$, $\bar{\omega}_{a} = \omega_{a}-\delta_\text{NM}$ and operators $\hat{\alpha},\hat{\beta}$ which eliminate the coupling term in Eq.~(\ref{eq:hamiltonian_a_b_int_anh}) whilst preserving canonical commutation relations
\begin{align}
\begin{split}
	\hat{H}/\hbar&=(\bar{\omega}_a+\lambda) \hat{\alpha}^\dagger\hat{\alpha}+\bar{\omega}_r \hat{\beta}^\dagger\hat{\beta}\\ 
	&\underbrace{- \frac{1}{12}\left(\chi_a^{1/4}\left(\hat{\alpha} + \hat{\alpha}^\dagger\right)+\chi_r^{1/4}\left(\hat{\beta}+\hat{\beta}^\dagger\right)\right)^4}_{\hat{H}_\text{anh}}\ .
	\label{eq:hamiltonian_a_b_anh}
\end{split}
\end{align}
The operators $\hat{\alpha},\hat{\beta}$ have a linear relation to $\hat{a},\hat{b}$, which determines the value of $\chi_a$ and $\chi_r$ (see App.~1). 
Expanding the anharmonicity leads to
\begin{align}
\begin{split}
	\hat{H}_\text{anh}/\hbar=&-\frac{\chi_{a}}{2}\left(\left(\hat{\alpha}^\dagger\hat{\alpha}\right)^2 +\hat{\alpha}^\dagger\hat{\alpha}+ \frac{1}{2}\right)\\
	&-\frac{\chi_{r}}{2}\left(\left(\hat{\beta}^\dagger\hat{\beta}\right)^2 +\hat{\beta}^\dagger\hat{\beta}+ \frac{1}{2}\right)\\
	&-2\chi_{ar}\left(\hat{\alpha}^\dagger\hat{\alpha}+\frac{1}{2}\right)\left(\hat{\beta}^\dagger\hat{\beta}+\frac{1}{2}\right)\ ,\\
	\label{eq:shifts}
\end{split}
\end{align}
if we neglect terms which do not preserve excitation number, irrelevant to first order in $\lambda$. This approximation is valid for $\lambda \ll |\Delta|,|3\omega_a-\omega_r|,|\omega_a-3\omega_r|$, which notably excludes the straddling regime~\cite{koch_charge-insensitive_2007}. 
The anharmonicity (or self-Kerr) of the normal-mode-splitted atom and resonator $\chi_a$ and $\chi_r$ is related to the AC Stark shift (or cross-Kerr) $2\chi_{ar}$ through
\begin{equation}
	\chi_{ar}=\sqrt{\chi_{a}\chi_{r}}\ .
\end{equation}
The AC Stark shift is the change in frequency one mode acquires as a function of the number of excitations in the other.

The appearance of an AC Stark shift and the resonators anharmonicity can be understood from the mechanism of normal-mode splitting. 
When the transmon and $LC$ oscillator dispersively couple, the normal-mode corresponding to the $LC$ oscillator will be composed of currents oscillating through its inductor but also partly through the JJ.
We can decompose the current $I$ traversing the JJ into the current corresponding to atomic excitations $I_a$ and resonator excitations $I_r$.
In Eq.~(\ref{eq:hamiltonian_a_b_anh}), this appears in the terms of flux as $\phi=\phi_a+\phi_r\propto \chi_a^{1/4}(\hat{\alpha} + \hat{\alpha}^\dagger)+\chi_r^{1/4}(\hat{\beta}+\hat{\beta}^\dagger)$.
Consequently the value of the JJ inductance is not only dependent on the number of excitations in the atom but also in the resonator.
Since the frequency of the normal-mode-splitted transmon and resonator depends on the value of this inductance, the atomic frequency is a function of the number of excitations in the resonator (AC Stark effect), and the resonator frequency changes as it is excited (the resonator acquires some anharmonicity).
Even when the resonator mode is in its ground state, vacuum current fluctuations shift the atomic frequency.
This can be verified by the presence of $1/2$ in the cross-Kerr term of Eq.~(\ref{eq:shifts}) which arise from commutation relations $[\hat{\alpha},\hat{\alpha}^\dagger]=[\hat{\beta},\hat{\beta}^\dagger]=1$, mathematically at the origin of vacuum fluctuations.

To summarize, compared to an isolated harmonic oscillator the energy levels of the coupled atom are shifted by: \textit{(1)} normal-mode splitting $\delta_\text{NM}$, \textit{(2)} its anharmonicity $\chi_a$ which arises from the quantum fluctuations of its eigen-states, and \textit{(3)} the shift proportional to $\chi_{ar}$ arising from the quantum fluctuations of the resonator it is coupled to.
These different effects are depicted in Fig.~\ref{fig:fig1}(b).
In Fig.~\ref{fig:fig2}(a,b), we show how these shifts manifest in a typical experimental setting where the detuning between the atom and resonator is varied, without explicitly showing contribution \textit{(2)}.
Off resonance, both modes are slightly shifted with respect to their un-coupled frequencies, and our theory allows us to distinguish the different effects which contribute to this shift.

\subsection{Analytical expression of the shifts in the RWA}

In the RWA $g\ll|\Delta|\ll\Sigma$, where $\Sigma = \omega_a +\omega_r$ the following approximations hold
\begin{align}
\begin{split}
	\bar{\omega}_a &= \omega_a- \delta_{NM} \simeq \omega_a- \frac{g^2}{\Delta} - \lambda\frac{g^2}{\Delta^2}\ ,\\
	\bar{\omega}_r &\simeq \omega_r+ \frac{g^2}{\Delta} + \lambda\frac{g^2}{\Delta^2}\ ,\\
	\chi_a &\simeq \lambda\left(1-2\frac{g^2}{\Delta^2}\right)\ ,\\
	\chi_r &= \mathcal{O}(g^4)\ ,\\
	\chi_{ar} &\simeq \lambda\frac{g^2 }{\Delta^2} ,
	\label{eq:shifts_rwa}
\end{split}
\end{align}
valid to leading order in $g$ and $\lambda$.
The expression for the AC Stark shift was also derived by Koch \textit{et al.}~\cite{koch_charge-insensitive_2007} from perturbation theory, given in the form $\lambda g^2/\Delta(\Delta-\lambda)$. 
Applying perturbation theory to the Hamiltonian of Eq.~(\ref{eq:hamiltonian_a_b_int_anh}), however, fails to predict the correct shift beyond the RWA and does not make the distinction between the physical origin of the different shifts.
Following Eqs.~(\ref{eq:shifts_rwa}), the total shift acquired when the resonator is in its ground-state $\delta\omega_a=\lambda-\delta_\text{NM}-\chi_a-\chi_{ar}$, is equal to $-g^2/\Delta$. 
This shift is equal to that of a harmonic oscillator coupled to another harmonic oscillator (here, the case $\lambda=0$) as well as that of a TLS coupled to a harmonic oscillator.
The fact that the total shift has the same magnitude in these three different systems can easily lead to a confusion as to its origin. 
In particular since the shift of a TLS is a purely quantum effect, whereas that of two coupled harmonic oscillators can be quantitatively derived from classical physics, and the weakly-anharmonic system lies somewhere in between.
This confusion can now be addressed: for a weakly-anharmonic system, there is a contribution from normal-mode splitting and from vacuum fluctuations which can both be quantified, and the former is much larger than the latter for a weakly-anharmonic system.
This also explains why earlier work~\cite{fragner_resolving_2008} found the Stark shift per photon to be smaller than the Lamb shift: vacuum fluctuations was not the only measured effect, normal-mode splitting also greatly contributed to the measured shift.
The proportion to which the total shift is due to vacuum fluctuations, as a function of anharmonicity, is shown in Fig.~\ref{fig:fig2}(c).

\subsection{Beyond the RWA}
% \label{sec:beyond}
\begin{figure}[t!]
\includegraphics[width=0.38\textwidth]{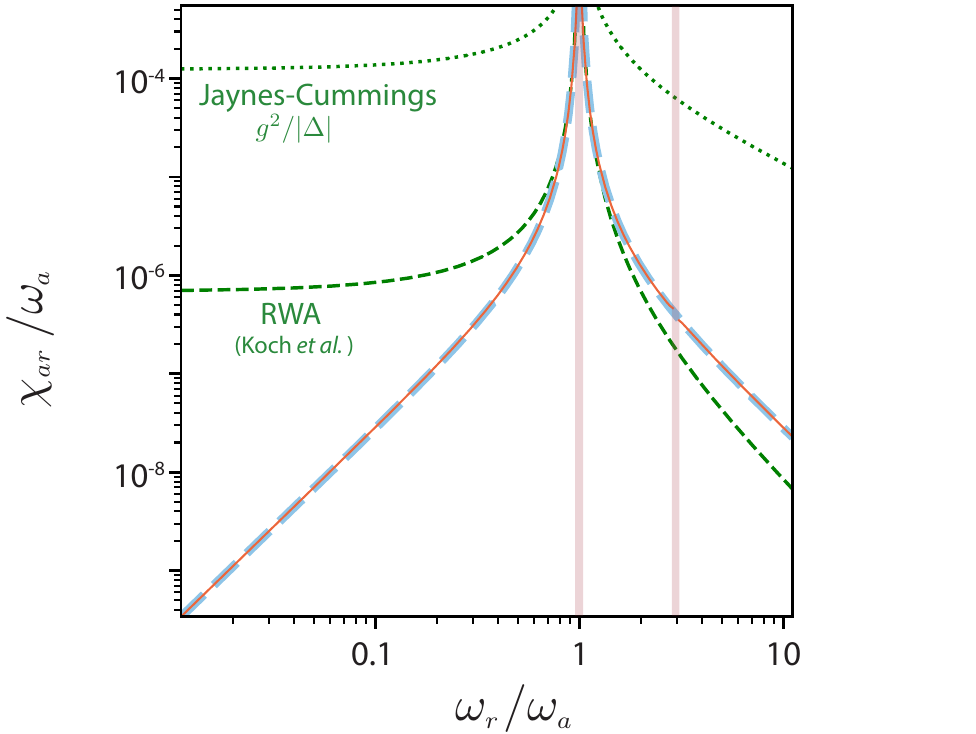}
\caption{Vacuum-fluctuations-induced shift $\chi_{ar}$ beyond the RWA fixing $\lambda/\omega_a=0.01$ and $g/\omega_a=0.02$.
Numerical calculation (full red line), are compared to the analytical expression of Eq.~(\ref{eq:shifts_beyond}) (dashed blue line) and Eq.~(\ref{eq:shifts}) (dashed green).
Resonances invalidating our approximations are denoted by red bars.}
\label{fig:fig3}
\end{figure}

Beyond the RWA to regimes of large detuning $g\ll|\Delta|\sim\Sigma$ the approximate expressions of the different shifts are given by

\begin{align}
\begin{split}
	\bar{\omega}_a &\simeq \omega_a- g^2\frac{2\omega_r}{\Delta\Sigma}- 4\lambda g^2\frac{\omega_r\omega_a }{\Delta^2\Sigma^2}\ ,\\
	\bar{\omega}_r &\simeq \omega_r+ g^2\frac{2\omega_a}{\Delta\Sigma}+ 4\lambda g^2\frac{\omega_a^2 }{\Delta^2\Sigma^2}\ ,\\
	\chi_a &\simeq \lambda\left(1-4g^2\frac{\omega_r \left(\omega_a^2 + \omega_r^2\right)}{\omega_a \Delta^2 \Sigma^2}\right)\ ,\\
	\chi_r &= \mathcal{O}(g^4)\ ,\\
	\chi_{ar} &\simeq 4\lambda g^2\frac{ \omega_r^2}{\Delta^2\Sigma^2} .
	\label{eq:shifts_beyond}
\end{split}
\end{align}
An important difference with the RWA is that the AC Stark shift $2\chi_{ar}$ scales with $\omega_r^2$, decreasing with the frequency of a coupled resonator as shown in Fig.~\ref{fig:fig3}.
This notably explains why the transmon is insensitive to low frequency charge fluctuations as compared to the highly anharmonic Cooper pair box.
It also explains why the transmon is not adapted to measuring individual quanta of far off-resonant systems such as low frequency mechanical oscillators~\cite{pirkkalainen_hybrid_2013}.
Contrary to the AC Stark shift in the RWA, this expression cannot be derived by applying perturbation theory to Eq.~(\ref{eq:hamiltonian_a_b_int_anh}).
The different shifts which arise from this method and perturbation theory are compared to two coupled harmonic oscillators and the two level system case in Supplementary Table~S1 and Fig.~S2~\cite{SI}.

\section{Summary and conclusions}

In conclusion, we presented a method to separate normal-mode splitting from the consequences of quantum fluctuations in the Hamiltonian of a weakly-anharmonic atom coupled to a harmonic oscillator. 
Through our theory, we reveal the physical origin of the different energy shifts arising in such a system. 
The main result is that only a small fraction of the total frequency shift can be attributed to quantum vacuum fluctuations, the dominant part being due to normal-mode splitting.
We prove that this small fraction can be experimentally measured as half the Stark shift per photon, for example in Ref.~\cite{fragner_resolving_2008}.
Extending this work to natural atoms (which are not perfect two-level systems either) also seems promising. 
Experiments in cavity QED show that the Lamb shift of natural atoms can be 40\% larger than half the Stark shift per photon~\cite{brune1994lamb}. 
As derived in this work, this indicates that the shift is not purely driven by quantum fluctuations.
Since the original picture of the Lamb shift is of a phenomenon driven by quantum fluctuations, our results raise questions about the terminology, and interpretation of, experiments in cavity and circuit QED.
In particular, should one reserve the terminology "Lamb shift" for only the part of the dispersive shift that arises from quantum fluctuations? 
In addition to addressing this fundamental question, we expect that the expressions derived in Eqs.~(\ref{eq:shifts_rwa}) and ~(\ref{eq:shifts_beyond}), as well as our approach to studying this Hamiltonian will become practical tools for experimental efforts in circuit QED. 

\section*{ACKNOWLEDGMENTS}

This project has received funding from the European Research Council (ERC) under the European Union’s Horizon 2020 research and innovation program (Grant Agreement No. 681476 - QOMD), and from the Netherlands Organisation for Scientific Research (NWO) in the Innovational Research Incentives Scheme – VIDI, project 680-47-526.

\appendix

\section*{Appendix: Transformation to the normal-mode basis}
The Hamiltonian 
\begin{equation}
    \hat{H}/\hbar=(\omega_a+\lambda) \hat{a}^\dagger\hat{a}+\omega_r \hat{b}^\dagger\hat{b}+g (\hat{a}-\hat{a}^\dagger)(\hat{b}-\hat{b}^\dagger)
    \label{eq:ham_no_bogo}
\end{equation}
describes two harmonic oscillators with a linear interaction between them. 
It can be compactly written as
\begin{align}
\begin{split}
  \hat{H}/\hbar&= \boldsymbol{v} ^T \boldsymbol{H} \boldsymbol{v}\ ,\\
\boldsymbol{v} ^T&= [\hat{a},\hat{b},\hat{a}^\dagger,\hat{b}^\dagger]\ ,\\
  \boldsymbol{H}= \frac{1}{2}&\begin{bmatrix}
 0 & g  & (\omega_a+\lambda) &-g     \\
 g & 0 &  -g &   \omega_r \\
 (\omega_a+\lambda) & -g  & 0 &g     \\
 -g & \omega_r &  g &   0 
\end{bmatrix}\ ,
    \label{eq:ham_no_bogo_matrix}
\end{split}
\end{align}
omitting constant contributions. Using this notation, the canonical commutation relations read
\begin{align}
\begin{split}
    [\boldsymbol{v},\boldsymbol{v} ^T]=\boldsymbol{v}\boldsymbol{v}^T-(\boldsymbol{v}\boldsymbol{v} ^T)^T=\boldsymbol{J}= \begin{bmatrix}
 0 & \boldsymbol{I}_2    \\
 -\boldsymbol{I}_2  & 0 \\\end{bmatrix}\ ,
\label{eq:commutation}
\end{split}
\end{align}
where $\boldsymbol{I}_2$ is the $2\times2$ identity matrix.
The objective of this section is to rewrite (\ref{eq:ham_no_bogo}) as the Hamiltonian of two independent harmonic oscillators, or normal-modes
\begin{equation}
    \hat{H}/\hbar=(\bar{\omega}_a+\lambda) \hat{\alpha}^\dagger\hat{\alpha}+\bar{\omega}_r \hat{\beta}^\dagger\hat{\beta}\ ,
    \label{eq:ham_after_bogo}
\end{equation}
which we write in compact notation as
\begin{align}
\begin{split}
    \hat{H}/\hbar&=\boldsymbol{\eta} ^T \boldsymbol{\Lambda} \boldsymbol{\eta}\ ,\\
    \boldsymbol{\eta} ^T&= [\hat{\alpha},\hat{\beta},\hat{\alpha}^\dagger,\hat{\beta}^\dagger]\\
    \boldsymbol{\Lambda}&=\frac{1}{2}\begin{bmatrix}
 0 & 0&(\bar{\omega}_a+\lambda)&0   \\
 0&0&0  & \bar{\omega}_r  \\
(\bar{\omega}_a+\lambda)&0&0  & 0 \\
 0&\bar{\omega}_r &0  & 0\\\end{bmatrix}\ .
    \label{eq:ham_bogo_matrix}
\end{split}
\end{align}
To do so, we need to find a matrix which maps $\boldsymbol{v}$ to a new set of annihilation and creation operators of the normal-modes $\boldsymbol{\eta}$ which should also satisfy the commutation relations (\ref{eq:commutation}). 

We start by noticing that the matrix $\boldsymbol{\Lambda}\boldsymbol{J}$ is diagonal
\begin{equation}
    \boldsymbol{\Lambda}\boldsymbol{J} = \frac{1}{2}\begin{bmatrix}
 -(\bar{\omega}_a+\lambda) & 0&0&0   \\
 0&-\bar{\omega}_r&0  & 0 \\
 0&0&(\bar{\omega}_a+\lambda)  & 0 \\
 0&0&0  & \bar{\omega}_r \\\end{bmatrix}\ ,
\end{equation}
and we define it as the diagonal form of the matrix $\boldsymbol{H}\boldsymbol{J}$. 
In other words, we can determine the value of $\bar{\omega}_a$ and $\bar{\omega}_r$ by diagonalizing $\boldsymbol{H}\boldsymbol{J}$.
An exact expression for these normal-mode frequencies is given by
\begin{align}
\begin{split}
&\bar{\omega}_{ar} = \frac{1}{\sqrt{2}}\bigg((\omega_a+\lambda)^2 + \omega_r^2 \\
	&\pm \sqrt{ ((\omega_a+\lambda)^2-\omega_r^2)^2 + 16 g^2 (\omega_a+\lambda) \omega_r }\bigg)^{\frac{1}{2}}\ .
\end{split}
\end{align}
As we will now demonstrate, defining $\boldsymbol{\Lambda}$ in this way will lead to operators with the correct commutation relations. 
We define the matrix of eigen-vectors that diagonalizes $\boldsymbol{H}\boldsymbol{J}$ as $\boldsymbol{F} = [\boldsymbol{w}_0,\boldsymbol{w}_1,\boldsymbol{w}_2,\boldsymbol{w}_3]$, such that
\begin{equation}
    \boldsymbol{H}\boldsymbol{J} = \boldsymbol{F}\boldsymbol{\Lambda}\boldsymbol{J}\boldsymbol{F}^{-1}
    \label{eq:eigen_vectors}
\end{equation}
The matrix $\boldsymbol{F}$ can be normalized in such a way that it satisfies an important condition, it can be made symplectic
\begin{equation}
    \boldsymbol{F}^T\boldsymbol{J}\boldsymbol{F}=\boldsymbol{F}\boldsymbol{J}\boldsymbol{F}^T=\boldsymbol{J}\ .
    \label{eq:sympletic}
 \end{equation} 
If the eigenvectors are normalized such that $\boldsymbol{w}_i^T \boldsymbol{w}_i=1$, the operation that leads to symplecticity is 
 \begin{align}
\begin{split}
\boldsymbol{w}_0' &= \pm\boldsymbol{w}_0/\sqrt{|\boldsymbol{w}_0^T\boldsymbol{J}\boldsymbol{w}_2|}\ ,\\
\boldsymbol{w}_1' &= \pm\boldsymbol{w}_1/\sqrt{|\boldsymbol{w}_1^T\boldsymbol{J}\boldsymbol{w}_3|}\ ,\\
\boldsymbol{w}_2' &= \pm\boldsymbol{w}_2/\sqrt{|\boldsymbol{w}_0^T\boldsymbol{J}\boldsymbol{w}_2|}\ ,\\
\boldsymbol{w}_3' &= \pm\boldsymbol{w}_3/\sqrt{|\boldsymbol{w}_1^T\boldsymbol{J}\boldsymbol{w}_3|}\ ,
\label{eq:normalization}
\end{split}
\end{align}
where the $+$ or $-$ sign is chosen such that if we redefine $\boldsymbol{F}= [\boldsymbol{w}_0',\boldsymbol{w}_1',\boldsymbol{w}_2',\boldsymbol{w}_3']$ it is of the form 
\begin{equation}
  \boldsymbol{F}= 
\begin{bmatrix}
 \boldsymbol{A}     & \boldsymbol{B}     \\
 \boldsymbol{B}     &   \boldsymbol{A}     \\
\end{bmatrix}\ ,
  \end{equation}
and such that $\boldsymbol{F}=\boldsymbol{I}_4$ in the limit $g=0$.
% We find that this method always produces a symplectic matrix although we currently have no rigorous proof. 
With $\boldsymbol{F}$ a symplectic matrix, we can define $\boldsymbol{\eta}$ as
\begin{equation}
    \boldsymbol{\eta} = \boldsymbol{F}^T \boldsymbol{v}\ 
    \label{eq:transformation}
\end{equation} 
and (\textbf{Proposition 1}) $\boldsymbol{\eta}$ will respect the commutation relations~(\ref{eq:commutation}) whilst ensuring that (\textbf{Proposition 2}) the two Hamiltonians (\ref{eq:ham_no_bogo_matrix}) and (\ref{eq:ham_bogo_matrix}) are equivalent. Proof of these proposition is provided at the end of this section. With the relation~(\ref{eq:identity}), we can invert~(\ref{eq:transformation}) to obtain
\begin{equation}
    \boldsymbol{v} = -\boldsymbol{J}\boldsymbol{F}\boldsymbol{J} \boldsymbol{\eta}\ .
    \label{eq:transformation2}
\end{equation} 
% with $+$ ($-$) corresponding to $\bar{\omega}_{r}$ ($\bar{\omega}_{a}$). 
Using the software Mathematica, we diagonalize $\boldsymbol{H}\boldsymbol{J}$ symbolically and perform the normalizations of Eqs.~(\ref{eq:normalization}) to obtain $\boldsymbol{F}$. 
As written in Eq.~\ref{eq:transformation2}, $\boldsymbol{F}$ leads to the transformation between the operators $\hat a$,$\hat b$ and $\hat\alpha$,$\hat\beta$. 
By Taylor expanding the resulting expressions for small values of $g$, we obtain
\begin{align}
\begin{split}
\hat{a} &\simeq \left( 1- g^2\frac{2  (\omega_a+\lambda) \omega_r}{\Delta'^2 \Sigma'^2}\right) \hat{\alpha}-\frac{g}{\Delta'}\hat{\beta} \\
&-g^2\frac{\omega_r}{(\omega_a+\lambda)}\frac{1}{\Sigma'\Delta'} \hat{\alpha}^\dagger- \frac{g}{\Sigma'}\hat{\beta}^\dagger\ ,  \\
\hat{b} &\simeq \frac{g}{\Delta'}  \hat{\alpha}+ \left(1- g^2\frac{2  (\omega_a+\lambda) \omega_r}{\Delta'^2 \Sigma'^2}\right) \hat{\beta} \\
&- \frac{g}{\Sigma'}  \hat{\alpha}^\dagger +g^2\frac{(\omega_a+\lambda)}{\omega_r}\frac{1}{\Sigma'\Delta'} \hat{\beta}^\dagger\ .\\
\end{split}
\end{align}
These approximations are valid to second order in $g$ and we define $\Delta'=\Delta-\lambda$ and $\Sigma'=\Sigma+\lambda$.
Using these relations, we can express the anharmonicity $\lambda (\hat{a}+\hat{a}^\dagger)/12$ as a function of $\hat{\alpha}$ and $\hat \beta$, leading to expressions for $\chi_a$ and $\chi_r$.
In the same approximation, the eigen-frequencies write
\begin{align}
\begin{split}
\bar{\omega}_{a} &\simeq\omega_a-\frac{2g^2\omega_r}{\Sigma'\Delta'}\ ,\\
\bar{\omega}_{r} &\simeq\omega_r+\frac{2g^2\omega_a}{\Sigma'\Delta'}\ .\\
\end{split}
\end{align}
leading to the expression for the normal mode splitting $\delta_\text{NM}$.
Finally, we provide proofs for the two propositions used above.

\textbf{Proposition 1:} this proof illustrates how essential it is that $\boldsymbol{F}$ be symplectic (Eq.~(\ref{eq:sympletic})) to obtain the desired commutation relations for $\hat\alpha$ and $\hat\beta$.
If $\boldsymbol{F}$ if symplectic, we find that the vector $\boldsymbol{\eta}$ satisfy the canonical commutation relations written in compact form in Eq.~\ref{eq:commutation}:

\begin{align}
\begin{split}
    [\boldsymbol{\eta},\boldsymbol{\eta} ^T]&=\boldsymbol{\eta}\boldsymbol{\eta}^T-(\boldsymbol{\eta}\boldsymbol{\eta} ^T)^T\\
        &\stackrel{\text{(\ref{eq:transformation})}}{=}\boldsymbol{F}^T (\boldsymbol{v}\boldsymbol{v}^T)\boldsymbol{F} - \boldsymbol{F}^T (\boldsymbol{v}\boldsymbol{v}^T)^T\boldsymbol{F}\\
        &=\boldsymbol{F}^T [\boldsymbol{v},\boldsymbol{v} ^T]\boldsymbol{F}\\
        &\stackrel{\text{(\ref{eq:commutation})}}{=}\boldsymbol{F}^T \boldsymbol{J}\boldsymbol{F} \\
        &\stackrel{\text{(\ref{eq:sympletic})}}{=}\boldsymbol{J}\ ,
\end{split}
\end{align}

\textbf{Proposition 2: } multiplying Eq.~(\ref{eq:sympletic}) with $\boldsymbol{J}$, we find
 \begin{equation}
    -\boldsymbol{F}\boldsymbol{J}\boldsymbol{F}^T\boldsymbol{J}=-\boldsymbol{J}\boldsymbol{F}\boldsymbol{J}\boldsymbol{F}^T=-\boldsymbol{J}^2=\boldsymbol{I}_4\ ,
    \label{eq:identity}
 \end{equation}
 where $\boldsymbol{I}_4$ is the $4\times4$ identity matrix.
 This relation allows us to introduce the matrix $\boldsymbol{F}$ into Eq.~(\ref{eq:ham_no_bogo_matrix})
 \begin{align}
 \begin{split}
 \hat{H}/\hbar&= \boldsymbol{v} ^T \boldsymbol{H} \boldsymbol{v}\\
        &\stackrel{\text{(\ref{eq:identity})}}{=}-\boldsymbol{v} ^T  \boldsymbol{H}\boldsymbol{J}\boldsymbol{F}\boldsymbol{J}\boldsymbol{F}^T \boldsymbol{v}\\
        &\stackrel{\text{(\ref{eq:eigen_vectors})}}{=}-\boldsymbol{v} ^T \boldsymbol{F}\boldsymbol{\Lambda}\boldsymbol{J}\underbrace{\boldsymbol{F}^{-1}\boldsymbol{F}}_{=\boldsymbol{I}_4}\boldsymbol{J}\boldsymbol{F}^T \boldsymbol{v}\\
        &\stackrel{\text{(\ref{eq:eigen_vectors})}}{=}-\boldsymbol{v} ^T \boldsymbol{F}\boldsymbol{\Lambda}\underbrace{\boldsymbol{J}\boldsymbol{J}}_{=-\boldsymbol{I}_4}\boldsymbol{F}^T \boldsymbol{v}\\
        &=(\boldsymbol{F}^T \boldsymbol{v})^T\boldsymbol{\Lambda}(\boldsymbol{F}^T \boldsymbol{v})\ ,
 \end{split}
 \end{align}
proving that $\hat{H}/\hbar=\boldsymbol{\eta}^T\boldsymbol{\Lambda}\boldsymbol{\eta}$.

\bibliography{library}

\end{document}